\documentclass[submission,copyright,creativecommons]{eptcs}
\providecommand{\event}{MEMICS 2016} 
\usepackage{breakurl}             
\usepackage{underscore}           
\usepackage{algorithm}
\usepackage{algpseudocode}
\usepackage{bm}
\usepackage{amsmath,amsthm}
\usepackage{enumerate}
\usepackage{multirow}
\usepackage{color}
\usepackage{graphicx}
\usepackage{array}

\definecolor{gr}{gray}{0.6}
\newcolumntype{C}[1]{>{\centering\let\newline\\\arraybackslash\hspace{0pt}}m{#1}}

\title{Using the Context of User Feedback in Recommender Systems}
\author{Ladislav Peska
\institute{Faculty of Mathematics and Physics }
\institute{Charles University in Prague \\ Malostranske namesti 25, Prague, Czech Republic}
\email{peska@ksi.mff.cuni.cz}
}
\def\titlerunning{Using the Context of User Feedback in Recommender Systemsr}
\def\authorrunning{Ladislav Peska}
\begin{document}
\maketitle

\begin{abstract}
Our work is generally focused on recommending for small or medium-sized e-commerce portals, where explicit feedback is absent and thus the usage of implicit feedback is necessary. Nonetheless, for some implicit feedback features, the  	\textit{presentation context} may be of high importance. In this paper, we present a model of relevant contextual features affecting user feedback, propose methods leveraging those features, publish a dataset of real e-commerce users containing multiple user feedback indicators as well as its context and finally present results of purchase prediction and recommendation experiments. 
Off-line experiments with real users of a Czech travel agency website corroborated the importance of leveraging  	\textit{presentation context} in both purchase prediction and recommendation tasks. 
\end{abstract}
\section{Introduction}
We face continuous growth of information on the web. The volume of products, services, offers or user-generated content rise every day and the amount of data on the web is virtually impossible to process directly by a human.  Automation of web content processing is necessary. Recommender systems aim to learn specific preferences of each distinct user and then present them surprising, unknown, but interesting and relevant items. Users do not have to specify their queries directly as in a search engine. Instead, their preferences are learned from their ratings (explicit feedback) or browsing behavior (implicit feedback). 

However, some domains, e.g., small or medium-sized e-commerce enterprises, introduce specific problems and obstacles making the deployment of recommender systems more challenging.  Let us list some of the obstacles:

\begin{itemize}
\item 	High concurrency has a negative impact on user loyalty. Typical sessions are very short, users quickly leave to other vendors, if their early experience is not satisfactory enough. Only a fraction of users ever returns. 
\item 	For those “single-time” visitors, it is not sensible to provide any unnecessary information such as ratings, reviews, registration details etc. 
\item  Consumption rate is low, users often visit only a handful of objects.
\end{itemize}

All the mentioned factors contribute to the data sparsity problem. Although the total number of users can be relatively large (hundreds or thousands per day), explicit feedback is very scarce and implicit feedback is also available only for a fraction of objects. Furthermore, as the space of potential implicit feedback features is quite large, it might be challenging to select the right approach to utilize them. In general, some rapidly learning algorithms, capable to recommend from only a limited feedback are needed.

Despite these obstacles, the potential benefit of deploying recommender systems is considerable, it can contribute towards better user experience, increased user loyalty and consumption and thus also improve vendor’s key success metrics.

Our work within this framework aims to bridge the data sparsity problem and the lack of relevant feedback by modelling and utilizing novel/enhanced sources of information, foremost implicit user feedback features.

More specifically, the work presented in this paper focuses on the question how to define and collect user preference 
\footnote{Please note that we will freely interchange \textit{user preference} and \textit{user engagement} concepts.} 
 in scenarios, where we cannot invasively ask users to provide it (i.e., there is no explicit feedback), but we can interfere with the website source code (and thus observe any type of user actions). 

\subsection{Main contributions}	
Main contributions of this paper are:
\begin{itemize}
\item 	Model of user feedback features enriched by the context of the page and device.
\item 	Methods interpreting this model of user feedback as a proxy of user engagement.
\item 	Experiments on real users of a Czech travel agency.
\end{itemize}
We also provide datasets of user feedback, contextual features and object’s attributes for the sake of repeatability and further experiments. 
\section{Related Work}
\subsection{Implicit Feedback in Recommender Systems}	
Contrary to explicit feedback, implicit feedback approach merely monitors user behavior without intruding it. Implicit feedback features varies from simple user visit or play counts to more sophisticated ones like scrolling or mouse movement tracking [5, 16]. Due to its effortlessness, data are obtained in much larger quantities for each user. On the other hand, they are inherently noisy and harder to interpret [4]. 

Our work lies a bit further from the mainstream of the implicit feedback research. To the best of our knowledge, vast majority of researchers focus on interpreting single type of implicit feedback, e.g., [17], or proposing various recommending algorithms while using predefined implicit feedback, e.g.,  [3, 4, 13, 14]. 

Our research goes towards modelling user’s preference and engagement based on multiple types of implicit feedback. We can trace such efforts also in the literature. One of the first paper mentioning implicit feedback was Claypool et al. [1] comparing several implicit preference indicators against explicit user rating. This paper was our original motivation to collect and analyze various types of user behavior to estimate user preference. More recently Yang et al. [16] analyzed several types of user behavior on YouTube. Authors described both positive and negative implicit indicators of preference and proposed a linear model to combine them. 

In our previous work, we defined a complex set of potentially relevant set of implicit user feedback features with respect to the e-commerce domain and provided software component collecting it [7]. We also show that using multiple types of feedback features provides significant improvements over using single feedback feature in purchase prediction task [9, 10]. However, in our previous works we used feedback features in its raw form without any respect to the context of the currently visited page or user’s browsing device, which can potentially affect user behavior.
\subsection{Context Awareness}	
In this paper, we focus on the  \textit{presentation context} (we will also refer to it as a \textit{context of page and device}) rather than more commonly utilized context of the user. We follow the hypothesis that if the same information is presented in a different form, the user’s response might differ as well. We can trace some notions of  \textit{presentation context} in the literature. For example, Radlinski et al. [12] and Fang et al. [3] considered object position as a relevant context for clickstream events. Also Eckhardt et al. [2] proposed to consider user ratings in the context of other objects available on the current page.

Closest to our work is the approach by Yi et al. [17], proposing to use  \textit{dwell time} as an indicator of user engagement. Authors discussed the role of several contextual features, e.g., content type, device type or article length on  	extit{dwell time} feedback. Nonetheless, there are several substantial differences between our approaches. First, Yi et al. focused solely on the  \textit{dwell time} and considered normalized  \textit{dwell time} directly as a proxy to the user engagement. Our approach is to integrate multiple indicators of user preference by using machine learning methods. Furthermore, the list of proposed contextual features are different as both the domains and data acquisition methods differ. We introduced, e.g., features based on page and browser window dimensions, not used in Yi et al. Last, Yi et al. proposed to utilize context merely to normalize \textit{dwell time}, however we include context in the feature engineering process.

\section{Materials and Methods}
\subsection{Outline of Our Approach}	

As already mentioned, the key part of our work aims on implicit user feedback, user preference and its usage in recommender systems. In traditional recommender systems, user $u$ rates some small sample $S$ of all objects $O$, which is commonly referred as user preference $r_{u,o}: o \in S \subset O$. The task of traditional recommender systems is to build suitable user model, capable to predict ratings $\hat{r}_{u,o'}$ of all objects $o' \in O$. If there are no explicit feedback, user preference must be inferred from implicit feedback. We denote this as inferred preference $\overline{r}_{u,o},o \in S$. If there is a single feedback feature $f_{u,o}$, the preference is usually inferred directly $f_{u,o} \approx\overline{r}_{u,o}$ [4]. However, some more elaborated approaches are necessary, if there are multiple feedback features $[f_1,...,f_i]$. 

Our approach is based on the hypothesis that purchases represent fully positive user preference:
\begin{equation} 
r_{u,o} :=\left\{
  \begin{array}{lr}
   1  \textrm{ IF }  u  \textrm{ bought }  o\\
    0  \textrm{ OTHERWISE }
  \end{array}
\right.
\end{equation} 
In another words, we promote  \textit{purchases} to the level of explicit user ratings. Unfortunately, the density of  \textit{purchases} is very low in the e-commerce
\footnote{Less than 0.4\% of the visited objects were purchased in our dataset.}
, so it is impractical to base recommendations directly on  \textit{purchases}. We also suppose that other visited, but not purchased objects reflect some level of user engagement, which can be inferred from other implicit feedback features. Our aim is to show that such inferred user preference provides better source information of a recommender system than binary visits or purchases. Our approach (see Fig. 1) is divided into three steps.

In the first step, feature engineering (Fig. 1c), we combined raw feedback features $F:[f_1,...,f_i]$, presentation context features $C:[c_1,...,c_j]$ and user statistics into a set of derived feedback features $\overline{F}:[\overline{f}_1,...,\overline{f}_i]$. Details of this procedure can be found in Section 3.2.

In the second step, the set of derived feedback features is transformed into the inferred user preference $[\overline{f}_1,...,\overline{f}_i]_{u,o} \to \overline{r}_{u,o}$ (Fig. 1d). The transformation is made via machine learning methods aiming to predict, whether the object $o$ was purchased by the user $u$, given the feedback $[\overline{f}_1,...,\overline{f}_i]_{u,o}$. More details can be found in Section 3.3.

Finally, we use $\overline{r}_{u,o}$ as an input of recommender systems to provide user with the list top-k objects (Fig. 1e). The description of used recommending algorithms can be found in Section 3.4. 

\begin{figure}[t!]
\includegraphics[width=160mm]{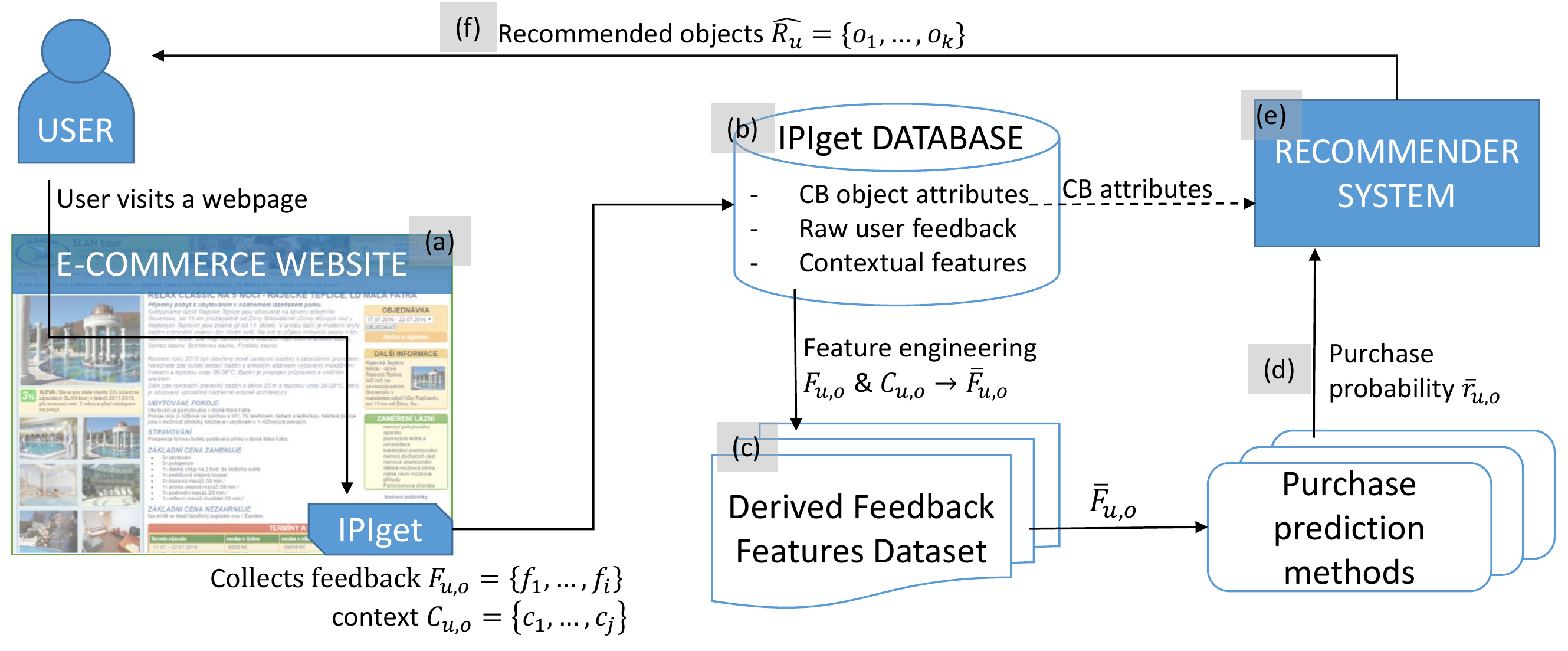}
\caption{Outline of our approach on utilizing complex user feedback and presentation context in recommender systems. Implicit feedback features and presentation context are collected by the IPIget tool (a) and stored in a database (b). Feature engineering process results into the set of derived feedback features (c) used for purchase prediction. The resulting purchase probability (d) serves as an input of a recommender system (e), which provides recommendations to the user (f). }
\label{fig:2}       
\end{figure}
\subsection{Implicit User Feedback and Presentation Context}	
In this section, we will describe the model of implicit feedback $F$, presentation context $C$ and feature engineering steps transforming it into derived feature set$ \overline{F}$. Raw feedback features and presentation context features are listed in Table 1 and Table 2. All features were collected with respect to the current user $u$ and object $o$.

\begin{table*}[t]
\centering
\caption{Description of the raw user feedback features.}
\label{tab:1}       
\begin{tabular}{ll}
\hline\noalign{\smallskip}
\textbf{\textit{Feature}} & \textbf{\textit{Description}} \\
\hline\noalign{\smallskip}
$f_1$	View Count &	The number of visits of the object \\
$f_2$	Dwell Time &	Total time spent on the object \\
$f_3$	Mouse Distance &	Approximate distance travelled by the mouse cursor \\
$f_4$	Mouse Time &	Total time, the mouse cursor was in motion \\
$f_5$	Scroll Distance &	Total scrolled distance \\
$f_6$	Scroll Time &	Total time, the user spent by scrolling \\
$ r$	\textbf{Purchase} &	Binary information whether user bought this object.  \\
\noalign{\smallskip}\hline
\end{tabular}
\end{table*}

\begin{table*}[t]
\centering
\caption{Description of the presentation context features.}
\label{tab:2}       
\begin{tabular}{ll}
\hline\noalign{\smallskip}
\textbf{\textit{Feature}} & \textbf{\textit{Description}} \\
\hline\noalign{\smallskip}
$c_1$	Number of links &	Total number of links presented on the page \\
$c_2$	Number of images &	Total number of images displayed on the page \\
$c_3$	Text size &	Total length of the text presented on the page \\
$c_4$	Page dimensions &	Width and height of the webpage \\
$c_5$	Browser window dimensions &	Width and height of the browser window \\
$c_6$	Visible area ratio &	Ratio between browser and page dimensions \\
$c_7$	Hand-held device &	Binary indicator, whether a cellphone or tablet is used \\
\noalign{\smallskip}\hline
\end{tabular}
\end{table*}

Let us now describe some features in more detail. Raw feedback features contain volumes of interaction generated by common user devices (mouse, keyboard etc.), or triggered by some GUI component. All the raw indicators have lower bounds equal to zero (i.e., no interaction was recorded) and except for purchases, they have no upper bound. We consider purchases as a golden standard for the user preference on e-commerce domains. 

Comparison between page and browser dimensions is crucial to determine necessity of scrolling the page and also serves as natural rate of the scrolled distance. Number of images, links, text and page sizes serve as a proxy to the page complexity, which should affect the volumes of user actions needed to fully evaluate the page. For example the page with higher amount of text usually takes longer time to read.

Derived feedback features were composed as follows. First, we defined relative per-user feedback features $f_i^u$ to be able to distinguish specific user’s browsing patterns (2), where $avg_u (f_i)$ denotes average of feature $f_i$ with respect to all records of user $u$.
\begin{equation}  	
f_i^u := f_i / avg_u (f_{i' } : u \textrm{ visited } i')	
\end{equation} 
Next, we defined two features, \textit{scrolled area} $f_{sc}$  and \textit{hit bottom page} $f_{hb}$, utilizing scrolling behavior and page dimensions. While the \textit{hit bottom} is a simple indicator, whether the page was fully scrolled, the \textit{scrolled area} represents the fraction of the page being visible for the user. Finally we aimed to relate volume of collected feedback with the page complexity context (\textit{number of links}, \textit{images} and \textit{text}, \textit{page dimensions} and \textit{visible area ratio}). As there is no single measure of the page complexity, we opted for the Cartesian product of feedback and reciprocal page complexity features $f_{i,j}$ (3). 
\begin{equation}  	
	f_{i,j}:= f_i / c_j ; \textrm{ where } f_i \in \{ f_1,...,f_6,f_1^u,..,f_6^u\}  \textrm{ and } c_j \in \{c_1, c_2, c_3, c_4, c_6 \} 
\end{equation} 
In our future work, we plan to investigate the experimental results with respect to the page complexity problem in order to deliver single page complexity metric. Table 3 lists all new features.

\begin{table*}[t]
\centering
\caption{Description of the features introduced in the feature engineering step.}
\label{tab:3}       
\begin{tabular}{l p{100mm}}
\hline\noalign{\smallskip}
\textbf{\textit{Feature}} & \textbf{\textit{Description}} \\
\hline\noalign{\smallskip}
$f_i^u$	Relative User Feedback &	The ratio between raw feedback and its per-user average value.\\
$f_{sc}$	Scrolled area &	The percentage of the page which have been presented in the browser visible area.\\
$f_{hb}$	Hit bottom & 	Binary indicator whether the user scrolled up to the bottom of the page.\\
$f_{i,j}$	Feedback vs. Page complexity &	The ratio between raw feedback (e.g., dwell time) and page complexity feature (e.g., number of links). \\
\noalign{\smallskip}\hline
\end{tabular}
\end{table*}

As our main aim is to evaluate contribution of presentation context to the recommendation quality, we defined and evaluated derived feedback datasets as follows:
\begin{itemize}
\item \textit{Dwell Time} dataset follows the recommendation from [17] on using dwell time as a proxy towards user engagement. It contains only features $f_2$ and $f_2^u$. 
\item \textit{Raw Feedback} dataset contains all $f_i$, $f_i^u$  feedback features but no context.
\item \textit{Raw + Context} dataset contains $f_i$, $f_i^u$, $c_i$, $f_{sc}$  and $f_{hb}$  features, but not $f_{i,j}$.
\item Finally, \textit{all features} dataset contains all described features ($f_i$, $f_i^u$, $c_i$, $f_{sc}$, $f_{hb}$ and $f_{i,j}$).
\end{itemize}
\subsection{Predicting Purchase Probability from User Feedback}	
In order to predict \textit{purchases} from derived feedback features, we have selected five machine learning techniques. For each technique, we used its R implementation from the caret package
\footnote{http://topepo.github.io/caret/}. As the \textit{purchase} indicator is a binary attribute, classification would be a natural option. However, our primary goal is not exactly predict purchased items. We need some better refined approximation for the user engagement as an input of the recommender system. Thus we need to focus either on classification method’s class probabilities, or consider purchase prediction as a regression task. We will further refer to both purchase probability (classification methods) and expected value of dependent variable (regression methods) as purchase probability $\overline{r}_{u,o}$. 

A potential advantage of regression techniques is the capability of providing negative preferences, i.e., infer user preference $<$ 0, but learning regression function from binary training data could be highly biased. Based on the previous discussion, we decided to evaluate following classification and regression methods in this task.

\textbf{\textit{Linear Regression (LinReg)}} is a simple regression method aiming to learn coefficients $\bm{A}$, $b$ with the minimal square loss of the linear function $y=\bm{AX}+b$, where $y$ is a dependent variable and $\bm{X}$ is a vector of independent variables ($r$ and $\overline{F}$ in our case).

\textbf{\textit{Lasso Regression (Lasso)}}: The least absolute shrinkage and selection operator is a regression method that performs feature selection, which makes it capable to deal with higher dimensional datasets. The LASSO’s objective is to find the parameter vector  $\bm{A}$ that minimizes the sum of squared errors plus the regularization term $\lambda \left\lVert \bm{A} \right\rVert_1$, where $\lambda$ is a hyperparameter controlling the regularization.

\textbf{\textit{AdaBoost regression (Ada LinReg)}}: Adaptive boosting is a meta-algorithm based on the principle of using weak learning algorithm iteratively over partially changed train sets. AdaBoost increases the weights of instances poorly predicted in previous iterations, thus although the individual learners are weak, the final model converge to a strong learner. In this case, linear regression was used.

\textbf{\textit{Decision tree classification (J48)}}: Specifically, the J48 implementation of the C4.5 algorithm was used. The C4.5 algorithm selects attributes on each node based on the normalized information gain. After the tree construction, it performs pruning, controlled by the hyperparameter $c$.

\textbf{\textit{AdaBoost classification (Ada Tree)}}: The algorithm is in principle the same as Ada LinReg, except that the decision stump was used as a weak learner in this case.
\subsection{Recommending based on Purchase Probability}	
The final step of our approach is to use purchase probability $\overline{r}_{u,o}$  in recommender systems. Our previous work [11] shown that purely collaborative algorithms are not suitable for small e-commerce enterprises, so we decided to evaluate one content-based and one hybrid recommending algorithm.

\textbf{\textit{Vector Space Model (VSM)}} is well-known content-based algorithm brought from information retrieval. We use the variant described in [6] with binarized content-based attributes serving as document vector, TF-IDF weighting and cosine similarity as objects’ similarity measure. The algorithm recommends top-k objects most similar to the user profile. 

For the purpose of content-based recommendation, the dataset of object’s (travel agency tours) attributes was used. The dataset contains approximately 20 attributes, such as type of the tour, accommodation quality, destination countries and regions, price per night, discount etc. For more information, please refer to [11].

\textbf{\textit{Popular from similar categories recommender (Popular SimCat)}}. Popular SimCat, is a simple hybrid approach based on collaborative similarity of product categories. There are two motivations for this algorithm. 

First, in our early experiments on a Travel Agency website [8], recommending objects from currently visited category turns out to be quite a good baseline. However, some categories were very narrow, containing only a handful of objects, sometimes even less than the intended size of the recommended objects list. For such a narrow category, it might be useful to also recommend objects from categories similar to the current one. Furthermore, there are substantially fewer categories than objects in the dataset (and the list of categories is much more stable), so it is possible to use collaborative similarity of categories.

Second, one of the most successful non-personalized recommendation approach is simply recommending the most popular objects. 

Putting both motivations together, the algorithm in training phase computes categories similarity and objects popularity: Categories similarity is defined as Jaccard similarity, based on the users covisiting both categories (4), where $U_{c1}$ and $U_{c2}$ are sets of users who visited category $c_1$ and $c_2$ respectively.

\begin{equation} 
Sim(c_1, c_2)  :=  \frac{|U_{c1} \cap U_{c2} |}{|U_{c1} \cup U_{c2}|}
\end{equation} 

The object’s popularity is defined as the logarithm of the number of object’s visits in the train set (5).

\begin{equation} 
Pop(o_i) := log \biggl(\sum_{\forall \textrm{ users}} {ViewCount(o_i)}\biggr) 
\end{equation} 

In the prediction phase, the algorithm collects all visited and similar categories for the current user and orders the objects according to the $Pop(o_i ) * Sim(c_{[o_i]})$ scoring function. More details can be found in [11].
\section{Evaluation and Results}
\subsection{Evaluation Protocol}
In this section, we would like to provide details of the evaluation procedure. In total four datasets of user feedback, five purchase prediction methods and two recommending algorithms were evaluated. Before we describe the protocol itself, let us mention some facts about the datasets used in the experiments. 

The dataset of user feedback (including contextual features) was collected by observing real visitors of a mid-sized Czech travel agency. The dataset was collected by the IPIget tool during the period of more than one year, contains over 560K records and is available for research purposes\footnote{See http://bit.ly/2dsjg6j}. For the purpose of the evaluation, we restricted the dataset only to the users, who visited at least 3 objects and purchased at least one of them. The resulting datasets contained 516 distinct users, 666 purchases, 1533 objects and over 23000 records, in average 45 records per user. However, please note that the number of records per user approximately follows the power-law distribution.

The evaluation of the proposed methods was carried out in two steps. 

In the first step, \textbf{\textit{purchase prediction}}, the task was to identify, which objects visited by the current user were purchased. Even though it looks like a binary classification, it is not exactly true, as we want a finer grained ordering as an input of the recommender system and we do not insist on proper classification of unpurchased items. We evaluate the problem as a ranking task, where ordering is induced by the purchase probability   $\overline{r}_{u,o}$. Objects actually purchased by the user should appear on top of the list. 

The evaluation was performed according to the leave-one-out cross-validation protocol applied on the user set. Machine learning algorithms were trained on the feedback data from all users, except for the current one, and afterwards predict for each object o visited by the current user u its purchase probability  $\overline{r}_{u,o}$. The ordering induced by  $\overline{r}_{u,o}$ was evaluated in terms of normalized discounted cumulative gain (nDCG), recall of purchased objects in top-k items (recall@top-k) and its average ranking position. 

This scenario simulates a well-known new user problem. When a new user enters the system, more complicated machine learning models cannot be retrained in real-time, taking into account feedback of the current user, so we need to infer his/her preferences from other users data. Using real-time local models, i.e., train only from the feedback of the current user, is impractical as there is usually not enough (if any) positive feedback.

The second step, \textbf{\textit{recommendation experiment}}, evaluates quality of the list of recommended objects in terms of position of the actually purchased ones. The evaluation of this step was also performed according to the leave-one-out cross-validation, however applied on the set of purchased objects. For each pair of the purchased object o and the user u who bought it, we trained recommender systems based on all other available data and ask it to recommend top-k best objects for the current user $\hat{R}_u :\{o_1,...,o_k\}$. Again, we consider the task as ranking, so the actually purchased object should appear on top of the list. Results were evaluated in terms of nDCG and recall@top-k metrics.
\subsection{Results: Purchase Prediction}
Table 4 depicts overall results of the purchase prediction experiment. The results of nDCG are surprisingly high, especially in case of \textit{Ada Tree} prediction method, however please note that the R implementation of nDCG metric\footnote{StatRank package, https://cran.r-project.org/web/packages/StatRank}  compensates for ties in the ranking. The results of other evaluation metrics (recall@top-k, average position) were very similar, so we omit them for the sake of space. Both classification methods clearly outperform all regression methods. Adding contextual features substantially improved prediction capability of all methods, but adding page complexity based features did not improve the results of all methods except for \textit{J48}. \textit{Ada Tree} classifier performed the best across all datasets.

\begin{table*}[t]
\centering
\caption{Results of the purchase prediction methods in terms of nDCG for different implicit feedback datasets. The best results are in bold.}
\label{tab:4}       
\begin{tabular}{lcccc}
\hline\noalign{\smallskip}
\textbf{\textit{Method}} & \textbf{\textit{DwellTime}} & \textbf{\textit{Raw feedback}} & \textbf{\textit{Raw + Context}} & \textbf{\textit{All feedback}} \\
\hline\noalign{\smallskip}
LinReg	 & 0.725	 & 0.714	 & 0.834	 & 0.828\\
Lasso	 & 0.730	 & 0.719	 & 0.831	 & 0.827\\
Ada LinReg	 & 0.713	 & 0.713	 & 0.863	 & 0.864\\
J48	 & 0.738	 & 0.740	 & 0.891	 & 0.893\\
Ada Tree	 & 0.757	 & 0.763	 & \textbf{0.950}	 &\textbf{0.950}\\
\noalign{\smallskip}\hline
\end{tabular}
\end{table*}

\begin{table*}[t]
\centering
\caption{Results of the recommendation experiment in terms of average nDCG for different implicit feedback datasets and recommending algorithms. Baseline methods are depicted in grey italics, the best results are in bold.}
\label{tab:5}       
\begin{tabular}{lccccc}
\hline\noalign{\smallskip}
\textbf{\textit{Method}} & \textbf{\textit{Recommender}} &\textbf{\textit{DwellTime}} & \textbf{\textit{Raw feedback}} & \textbf{\textit{Raw + Context}} & \textbf{\textit{All feedback}} \\
\hline\noalign{\smallskip}
\textit{\color{gr}Binary }& 	\textit{\color{gr}VSM} & 	\multicolumn{4}{c}{\textit{\color{gr}0.304}}\\
LinReg & 	VSM & 	0.299 & 	0.297 & 	0.215 & 	0.215\\
Lasso & 	VSM & 	0.304	 & 0.298 & 	0.213 & 	0.216\\
Ada LinReg & 	VSM & 	0.302 & 	0.301 & 	0.215 & 	0.215\\
J48 & 	VSM & 	0.299	 & 0.295 & 	0.303	 & \textbf{0.311}\\
Ada Tree & 	VSM	 & 0.293	 & 0.298	 & 0.294 & 	0.296\\
\textit{\color{gr}Binary	 }& \textit{\color{gr}Popular SimCat} & 	\multicolumn{4}{c}{\textit{\color{gr}0.362}}\\
LinReg & 	Popular SimCat & 	0.342	 & 0.342	 & 0.267	 & 0.270\\
Lasso	 & Popular SimCat	 & 0.359	 & 0.343	 & 0.260	 & 0.270\\
Ada LinReg	 & Popular SimCat & 	0.361	 & 0.360	 & 0.264	 & 0.264\\
J48	 & Popular SimCat	 & 0.353	 & 0.354	 & \textbf{0.373} & 	0.372\\
Ada Tree	 & Popular SimCat	 & 0.358	 & 0.358 & 	0.363	 & 0.370\\
\noalign{\smallskip}\hline
\end{tabular}
\end{table*}
\subsection{Results: Recommendation Experiment}
Table 5 depicts the overall results of the recommendation experiment. Additionally, to the purchase probability inputs we also evaluated  \textit{Binary} baseline method, which simply considers all visited objects as relevant\footnote{We did not evaluate the input based solely on purchases, because over 90\% of users purchased only one item and recommending algorithms could not predict anything for them.}. For the sake of space, we do not display detailed the results of recall@top-k metric, however it mostly corresponds with nDCG. The best performing method in terms of recall@top-5 and recall@top-10 was  \textit{J48} with  \textit{Raw+Context} dataset and  \textit{Popular SimCat} recommender, achieving recall of 0.297 and 0.376 for top-5 and top-10 respectively.

As can be seen from the results, all regression based methods performed worse than the baseline and furthermore its performance gradually decreased for enriched datasets in the most cases. This might be a problem of learning regression from only binary input, but as all regression methods were based on a linear model, we do not want to conclude on this subject yet. On the other hand, \textit{Popular SimCat} with both \textit{Ada Tree} and \textit{J48}, as well as \textit{VSM} with \textit{J48} outperformed baselines. Furthermore as can be seen in Table 6, there is a significant performance improvement between \textit{raw feedback} and datasets containing contextual features for those methods. It seems that using page complexity based features $f_{i,j}$ can also improve performance of some methods, however the results are less clear at this point.

Surprisingly, the relatively simple \textit{Popular SimCat} algorithm produced consistently better results than \textit{VSM}. This is in contradiction with our previous experiments with these algorithms [11], however we need to note that the target of the previous experiment was to predict visited instead of purchased objects. We would like to investigate this topic more in our future work.

\begin{table*}[t]
\centering
\caption{P-values of the binomial significance test [15] for selected combination of algorithms. The test was performed with respect to the recall of purchased object in top-K.}
\label{tab:6}       
\begin{tabular}{lllC{15mm}C{15mm}}
\hline\noalign{\smallskip}
\textbf{\textit{Recommender}} & \textbf{\textit{Baseline}} &\textbf{\textit{Method}} & \textbf{\textit{p-value}} \newline  recall@5 & \textbf{\textit{p-value}}\newline  recall@10\\
\hline\noalign{\smallskip}
\multirow{2}{*}{VSM}	 & \textit{\color{gr}Binary }	 & J48 (All feedback) & 	0.028	 & 0.026\\
	 &J48 (Dwell time) & 	J48 (All feedback) & 	0.024 & 	0.001\\
\hline
\multirow{5}{*}{Popular SimCat}	 & \textit{\color{gr}Binary }	 & J48 (All feedback) & 	0.025 & 	0.198\\
	 &\textit{\color{gr}Binary }	 & J48 (Raw + Context) & 	0.036 & 	0.015\\
	 &\textit{\color{gr}Binary }	 & Ada Tree (All feedback) & 	0.009	 & 0.154\\
	 &J48 (Raw feedback) & 	J48 (All feedback) & 	0.001	 & 0.004\\
	 &Ada Tree (Raw feedback) & 	Ada Tree (All feedback) & 	0.057	 & 0.000\\
\noalign{\smallskip}\hline
\end{tabular}
\end{table*}

\section{Conclusions and Future Work}
In this paper, our aim was to show that user feedback should be considered with respect to the context of the page and device. We defined several features describing such context and incorporate them into the user feedback feature space. In the purchase prediction task, the usage of context clearly improved performance of all learning methods in predicting purchased objects. Furthermore, by using purchase probability as a proxy towards user engagement, we were able to improve quality of the recommendations over both binary feedback baseline and uncontextualized feedback in terms of nDCG and recall@top-k. 

In this paper we did not investigate the influence of each contextual feature separately as well as possibility to combine purchase probabilities coming from different learning methods. Both should be done in our future work. The presented approach can be applied on any domain, as long as there is some natural indicator of user engagement or preference (like purchases in e-commerce). Thus, naturally, one possible direction of our research is to extend this approach beyond its current e-commerce application. 

Another task is to combine the contextual approach with our previous work, e.g., on using early user feedback on lists of objects [11] and corroborate the results in on-line experiments.

\textbf{Acknowledgments}. The experiments presented in this paper were done while author was a Ph.D. student at Charles University in Prague. The work was supported by the Czech grant P46. Supplementary materials (datasets, source codes, results) can be found on: http://bit.ly/2dsjg6j.

\nocite{*}
\bibliographystyle{eptcs}
\bibliography{generic}
\end{document}